\documentclass[twocolumn,showpacs,preprintnumbers,superscriptaddress,nofootinbib]{revtex4}
\usepackage{graphicx}

\def\lsim{\mathrel{\raise.3ex\hbox{$<$\kern-.75em\lower1ex\hbox{$\sim$}}}}
\def\gsim{\mathrel{\raise.3ex\hbox{$>$\kern-.75em\lower1ex\hbox{$\sim$}}}}

\def\eps{\epsilon}
\def\eplus{e^+}
\def\neubar{\bar{\nu}}
\def\eminusi{e^-_i} 
\def\eminusf{e^-_f}
\def\shat{\hat{s}}
\newcommand{\nc}{\newcommand}
\nc{\beq}{\begin{equation}}   \nc{\eeq}{\end{equation}}
\nc{\bea}{\begin{eqnarray}}   \nc{\eea}{\end{eqnarray}}
\newcommand{\gvdl}{g_V^{dL}}
\newcommand{\gvdr}{g_V^{dR}}
\newcommand{\gvul}{g_V^{uL}}
\newcommand{\gvur}{g_V^{uR}}
\newcommand{\yl}{y_{Htb}^L}
\newcommand{\yr}{y_{Htb}^R}

\begin{document}

\preprint{CALT-68-2448}
\preprint{MADPH-03-1342}
\preprint{UCD-03-09}
\preprint{hep-ph/0308124}

\title{$e^+e^- \to H^+ e^- \bar \nu$ in the two-Higgs-doublet model}

\author{Tom Farris}
\email{farris@physics.ucdavis.edu}
\affiliation{Davis Institute for High Energy Physics,
University of California, Davis, CA 95616, USA}

\author{Heather E. Logan}
\email{logan@pheno.physics.wisc.edu}
\affiliation{Department of Physics, University of Wisconsin,
Madison, Wisconsin 53706, USA} 

\author{Shufang Su}
\email{shufang@theory.caltech.edu}
\affiliation{California Institute of Technology, Pasadena, 
California 91125, USA}
\affiliation{Department of Physics, University of Arizona, Tucson, 
Arizona 85721, USA}

\begin{abstract}
We calculate the cross section for $e^+e^- \to H^+ e^- \bar \nu$ in the
two-Higgs-doublet model from one-loop diagrams involving top and bottom 
quarks.  
This process offers the possibility of producing the charged Higgs boson 
at the $e^+e^-$ collider when its mass is more than half the center-of-mass 
energy, so that charged Higgs pair production is kinematically forbidden.
The cross section receives contributions from both $s$-channel
and $t$-channel processes; the $s$-channel contribution dominates
for center-of-mass energies of 1 TeV and below.  About 80\% of the 
$s$-channel contribution comes from the resonant process
$e^+e^- \to H^+ W^-$, with $W^- \to e^- \bar\nu$.
The cross section is generally small, below 0.01 fb for 
$\tan\beta > 2$, and falls with increasing $\tan\beta$.
\end{abstract}

\date{January 12, 2004}


\maketitle


\section{\label{sec:intro}Introduction}

The Higgs mechanism provides an elegant way to explain electroweak symmetry
breaking and the origin of the masses of the Standard Model fermions.
In the Standard Model with a single Higgs doublet, however, the mass of the 
Higgs boson (and, therefore, the energy scale of electroweak symmetry
breaking) is quadratically sensitive to physics at high energy scales
via radiative corrections.  This sensitivity leads to a fine-tuning problem
between the electroweak scale and the cutoff (grand unification or Planck) 
scale.

Models that address the fine-tuning problem often enlarge the Higgs sector.
For example, the Higgs sectors of the 
Minimal Supersymmetric Standard Model (MSSM) \cite{HaberKane},
topcolor-assisted technicolor \cite{TC2}, 
and some of the ``little Higgs'' models \cite{LH2HDM}
contain two Higgs doublets with electroweak-scale masses.
This enlargement of the Higgs sector leads to the presence of a charged
Higgs boson in the physical spectrum, in addition to extra
neutral states.  Observation of these extra Higgs bosons and the measurement
of their properties is a central goal in searches for physics beyond the 
Standard Model.  In this paper we focus on the charged Higgs boson, $H^{\pm}$.

Unlike the CP-even Higgs sector, in which at least one of 
the CP-even Higgs bosons is guaranteed to be detected at future 
colliders \cite{lhcATLAS, lhcCMS, perini, Espinosa:1998xj}, the 
discovery of the heavy charged Higgs boson poses a special 
experimental challenge.
Studies of charged Higgs boson production at present and future colliders
are generally done in the context of the MSSM, or more generically, in a
general two Higgs doublet model (2HDM). 
Searches for $H^+H^-$ pair production at the CERN LEP-2 collider
only set a lower bound on the charged Higgs mass of $m_{H^{\pm}} > 78.6$ GeV
\cite{LEPHpm}. 
At Run II of the Fermilab Tevatron,
now in progress, the charged Higgs boson could be discovered in top 
quark decays if $m_{H^{\pm}} \lsim m_t$ and if $\tan\beta$ (the ratio 
of vacuum expectation values of the two Higgs doublets) is large
\cite{topdecays}.  No sensitivity is expected in direct production 
unless QCD and supersymmetry (SUSY) effects conspire to enhance the 
cross section \cite{BelyaevHpm}.

The difficulty of discovering a heavy charged Higgs boson
comes from the fact that there are no tree level 
$W^\pm Z H^\mp$ and $W^\pm \gamma H^\mp$ couplings. 
This forbids potential tree-level
discovery processes like $W^{\pm*}\rightarrow H^{\pm}Z, H^{\pm}\gamma$ and 
$Z^*\rightarrow H^{\pm} W^{\mp}$ at the Tevatron, weak boson fusion 
$W^{\pm*}Z^*\rightarrow H^{\pm}$ at the 
CERN Large Hadron Collider (LHC), and 
$e^+e^- \rightarrow Z^* \rightarrow H^{\pm}  W^{\mp}$, 
$e^+e^- \rightarrow  H^+ e^- \bar{\nu}$ at a future linear $e^+e^-$ 
collider (LC).
In addition, the charged Higgs can not be produced via the $s$-channel 
gluon fusion process at the LHC.  

Charged Higgs bosons can of course be pair produced at tree level.  At hadron 
colliders, however, the coupling is electroweak in strength which 
leads to a small production cross section so that the signal has a hard time
competing with 
the huge QCD background.  At the LC, the charged Higgs
boson can be pair produced via $e^+e^- \to H^+H^-$.  This
process requires $m_{H^{\pm}} < \sqrt{s}/2$, where $\sqrt{s}$
is the LC center-of-mass energy.  This process is thus
only useful for a relatively light charged Higgs boson; 
for an experimental study, see Ref.~\cite{Kiiskinen}.
For $m_{H^{\pm}} > \sqrt{s}/2$, the charged Higgs boson must be
produced singly.

Let us consider the production of a single charged Higgs boson at tree level
(without any other heavy Higgs bosons in the final state).  
The only relevant process is a charged Higgs boson produced together 
with third generation quarks or leptons.  At the LHC, $gb \to H^- {t}$ 
with $H^- \to \tau^- \bar\nu$ is only good for large $\tan\beta$, where 
the bottom and tau Yukawa couplings are enhanced.  
This discovery channel will cover 
$\tan\beta \gsim 10$ for $m_{H^{\pm}} = 250$ GeV
($\tan \beta \gsim 17$ for $m_{H^{\pm}} = 500$ GeV)
\cite{LHCHpm}.
Within the context of the MSSM, the absence of a neutral Higgs boson 
discovery at LEP-2 excludes the range $0.5 < \tan\beta < 2.4$ at 95\%
confidence level \cite{LEP2}.  (In a general 2HDM, however, these low
values of $\tan\beta$ are not yet excluded.)
This leaves a wedge-shaped region of parameter
space at moderate $\tan\beta$ in which the charged Higgs boson 
would not be discovered at the LHC.
At the LC, the processes 
$e^+e^-\rightarrow H^+ \bar{t}b, H^+\tau^-\bar{\nu}$ have been 
considered in the literature \cite{tbHtaunuH}.  
At a 500 GeV LC with integrated 
luminosity of 500 ${\rm fb}^{-1}$, both $H^+ \tau^- \bar \nu $ and 
$H^+ \bar t b$ production yield 
$\geq 10$ events at large $\tan\beta \sim 40$ for $m_{H^{\pm}} \lsim 270$ GeV,
with $H^+ \tau^- \bar \nu $ production having a slightly larger cross
section \cite{tbHtaunuH}.
At a 1000 GeV LC with integrated 
luminosity of 1000 ${\rm fb}^{-1}$, $\bar t b H^+$ 
production is more promising, due to
the larger phase space available; it yields $\geq 10$ events at large
$\tan\beta \sim 40$ for $m_{H^{\pm}} \lsim 550$ GeV, while 
$\tau^- \bar \nu H^+$ production gives a reach of only $m_{H^{\pm}} \lsim 520$
GeV \cite{tbHtaunuH}. 
At low $\tan\beta = 1.5$, the reach in the $H^+ \bar t b$ channel is the 
same as at $\tan\beta = 40$.

The charged Higgs boson can also be produced together with light SM particles 
via loop induced processes, e.g., $e^+e^-\rightarrow H^{\pm}W^{\mp}$
and $e^+e^-\rightarrow H^+ e^-\bar\nu$.
The first process, $e^+e^- \to H^+W^-$, was computed in the general
2HDM in Refs.~\cite{Arhrib,Kanemura,Zhu}.  The additional one-loop
diagrams involving SUSY particles were computed in Refs.~\cite{HW, brein},
and the behavior of the cross section as a function of the SUSY parameter
space was explored in Ref.~\cite{HWupdate}.
This process could extend the reach in $m_{H^{\pm}}$ at low $\tan\beta$
beyond the pair production threshold. At a 500 GeV LC with integrated 
luminosity of 500 ${\rm fb}^{-1}$, more than ten events can be 
produced in this channel for $m_{H^{\pm}}$ up to 330 GeV and $\tan\beta$ up
to 4.7 in the 2HDM, while using an $80\%$ left-polarized electron beam 
or including
contributions from light superpartners can increase the cross section 
further.

In this paper, we expand upon these earlier results by computing
the cross section for the process $e^+e^- \to H^+e^-\bar\nu$ in the 
Type II 2HDM.
In addition to the $s$-channel process 
$e^+e^- \to H^+W^-$ in which the $W^-$ decays
to $e^- \bar \nu$, this process also receives contributions from 
$t$-channel gauge boson exchange.  The $t$-channel diagrams are 
potentially much more significant than $s$-channel diagrams at high 
collider center-of-mass energies.  Finally, our approach also
takes into account non-resonant contributions
in the $s$-channel process.

In our calculation we include only the one-loop 
diagrams involving top and bottom 
quarks.  We neglect diagrams involving gauge and/or Higgs
bosons in the loop.
It was shown in Ref.~\cite{Kanemura} for the $e^+e^- \to H^+W^-$ process that
the diagrams involving gauge and Higgs bosons are negligible unless
the Higgs boson self-couplings are large.  In the MSSM, the Higgs boson
self-couplings are related to gauge couplings and are relatively small, 
so that the gauge and Higgs boson loops are not important.  In a general
2HDM with large but perturbative Higgs boson self-couplings, the gauge
and Higgs boson loop contributions to the $e^+e^- \to H^+W^-$ cross 
section can be an order of magnitude larger than the top and bottom quark 
loops, but only for moderate to large values of $\tan\beta$ where the
cross section is already quite small \cite{Kanemura}.
In the MSSM, the process $e^+e^- \to H^+e^-\bar\nu$ will also get contributions
from loops involving SUSY particles.  Their calculation is beyond the scope
of our current work\footnote{Preliminary results on the 
cross section of $e^+e^- \to H^+e^-\bar\nu$ in the framework of MSSM 
were shown in \cite{sventalk}.}.
However, we can estimate their impact based on the results of
Refs.~\cite{HW,HWupdate}; as we will show, for $\sqrt{s} \lsim 1$ TeV,
the $e^+e^- \to H^+e^-\bar\nu$ cross section is dominated by the 
$s$-channel contribution, which is in turn dominated by the resonant
$e^+e^- \to H^+W^-$, $W^- \to e^- \bar\nu$ contribution.  This cross
section can be enhanced by up to 50-100\% at low $\tan\beta$ by loops
involving light superpartners \cite{HW,HWupdate}.

This paper is organized as follows.  In Sec.~\ref{sec:calculation} we
lay out the formalism for the calculation and define form-factors for
the one-loop $W^+H^-Z$ and $W^+H^-\gamma$ couplings.  
We also discuss the renormalization procedure
and define the counterterms.  In Sec.~\ref{sec:numerics} we present
our numerical results.  In Sec.~\ref{sec:modeldependence} we discuss
the extension of our results to the Type I 2HDM, larger extended Higgs
sectors, and the MSSM.
Sec.~\ref{sec:conclusions} is reserved for our
conclusions.  We summarize our notation and conventions 
in Appendix~\ref{sec:notation} and give the full expressions for the 
matrix elements and their squares in Appendix~\ref{sec:matrixelements}.
A derivation of the sum of the contributions from $H^+W^+$ and $H^+G^+$
mixing is given in Appendix~\ref{sec:sum}.

\section{\label{sec:calculation}Calculation}

Top and bottom quark loops contribute to the process 
$e^+e^- \to H^+e^-\bar\nu$ by inducing an effective 
$W^+VH^-$ coupling (with $V = \gamma,Z$) and by generating mixing
between $H^+$ and the $W^+$ and $G^+$ bosons, which must be renormalized.  
We work in the 't Hooft-Feynman gauge, in which $G^+$ is the Goldstone 
boson that corresponds to the longitudinal component of $W^+$.

\subsection{\label{sec:formalism}Form-factors}

\begin{figure}
\resizebox{8.5cm}{!}{\includegraphics*[25,570][350,690]{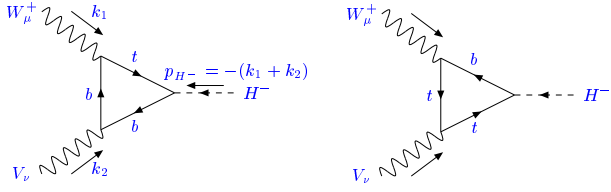}}
\caption{Top and bottom quark contributions to the one-loop
$W^{+ \mu}V^{\nu}H^-$ vertex, where $V=\gamma,Z$.}
\label{fig:tb}
\end{figure}

We define the effective $W^{+ \mu}(k_1)V^{\nu}(k_2)H^-$ coupling 
as follows (with all particles and momenta incoming as shown 
in Fig.~\ref{fig:tb}):
        \begin{equation}
        i \mathcal{M}^{\mu\nu} = i \left[ G_V g^{\mu\nu}
        + H_V k_1^{\nu} k_2^{\mu} 
        + F_V i \epsilon^{\mu\nu\alpha\beta} k_{1 \alpha} k_{2 \beta} \right].
        \label{eq:formfactors}
        \end{equation}
Here $k_1$ is the incoming momentum of the $W^+$, $k_2$ is the incoming
momentum of $V = \gamma,Z$, and $\epsilon^{0123}=1$.
Assuming CP conservation,
the effective coupling for the Hermitian conjugate vertex 
$W^{- \mu}(k_1)V^{\nu}(k_2)H^+$ is given by Eq.~(\ref{eq:formfactors})
with $F_V \to -F_V$.
The diagrams involving top and bottom quarks are shown in Fig.~\ref{fig:tb}.
Explicit expressions for the form factors $G_V$, $H_V$ and $F_V$ 
are given in Appendix~\ref{sec:matrixelements}.

\begin{figure}
\resizebox{8.5cm}{!}{\includegraphics*[0,560][360,700]{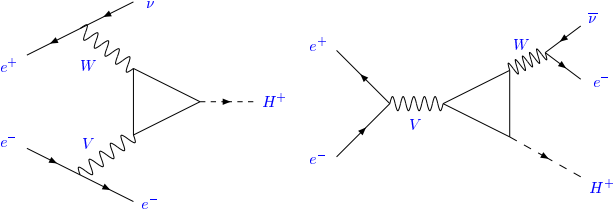}}
\caption{One-particle-irreducible contributions to 
$e^+e^- \to H^+ e^- \bar \nu$ due to the fermion triangle diagram
shown in Fig.~\ref{fig:tb}.}
\label{fig:1PI}
\end{figure}

The effective $W^+VH^-$ coupling
gives both $t$-channel and $s$-channel
contributions to the matrix element for $e^+e^- \to H^+ e^- \bar \nu$,
as shown in Fig.~\ref{fig:1PI}.  
The corresponding matrix elements can be written as
        \begin{eqnarray}
        i \mathcal{M}_{\rm loop}^t &=& 
        i e^2 g_W
        \left[ G_V g^{\mu\nu} + H_V k_1^{t\nu} k_2^{t\mu}
        + F_V i \epsilon^{\mu\nu\alpha\beta} k_{1\alpha}^t k_{2\beta}^t \right]
        \nonumber \\
        &&\hspace{-0.5 in}\times
        \frac{\bar v(e^+_i) \gamma_{\mu} P_L v(\bar\nu_f)
        \bar u(e^-_f) \gamma_{\nu}(g^{eL}_V P_L + g^{eR}_V P_R) u(e^-_i)}
        {[k_1^{t2} - m_W^2][k_2^{t2} - m_V^2]}  
        \label{eq:Mformfactors_t} \\
        i \mathcal{M}_{\rm loop}^s &=&
        i e^2 g_W
        \left[ G_V g^{\mu\nu} + H_V k_1^{s\nu} k_2^{s\mu}
        + F_V i \epsilon^{\mu\nu\alpha\beta} k_{1\alpha}^s k_{2\beta}^s 
        \right]
        \nonumber \\
        &&\hspace{-0.7 in}\times
        \frac{\bar u(e^-_f) \gamma_{\mu} P_L v(\bar\nu_f)
        \bar v(e^+_i) \gamma_{\nu} (g_V^{eL} P_L + g_V^{eR} P_R) u(e^-_i)}
        {[k_1^{s2} - m_W^2 + im_W\Gamma_W][k_2^{s2} - m_V^2]},
        \label{eq:Mformfactors_s}
        \end{eqnarray}
where we use the notation $e^-_i \equiv p_{e^-_i}$, etc.  
The momenta $k_{1,2}^{t,s}$ are given by:
        \begin{eqnarray}
        &&k_1^t = e^+_i - \bar \nu_f, \qquad 
        k_2^t = e^-_i - e^-_f, \nonumber \\
        &&k_1^s = - e^-_f - \bar\nu_f, \qquad
        k_2^s = e^-_i + e^+_i.
        \end{eqnarray}
$P_{R,L} = (1 \pm \gamma^5)/2$ are the right- and 
left-handed projection
operators.  The gauge couplings $g_W$ and $g^{eL,R}_V$ are given in 
Appendix~\ref{sec:notation}.
The additional contribution from the counterterm for the 
$W^+VH^-$ vertex is given
in the next subsection.

\subsection{\label{sec:renormalization}Renormalization}

\begin{figure}
\resizebox{8.5cm}{!}{\includegraphics*[55,455][340,740]{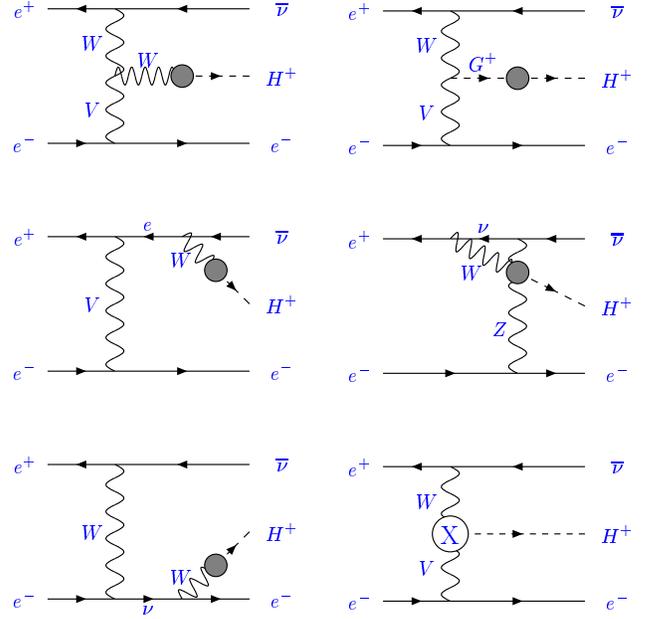}}
\caption{The $t$-channel mixing self-energy and counterterm 
contributions to $e^+e^- \to H^+ e^- \bar \nu$.  The shaded blob
denotes the renormalized mixing self-energy and the $X$ denotes the
$W^{+\mu}V^{\nu}H^-$ counterterm.}
\label{fig:rent}
\end{figure}

\begin{figure}
\resizebox{8.5cm}{!}{\includegraphics*[50,450][365,750]{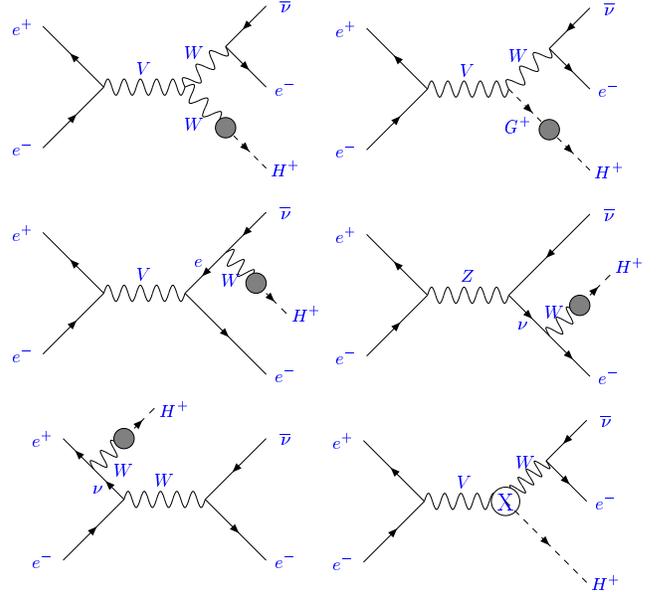}}
\caption{As in Fig.~\ref{fig:rent} but for the $s$-channel contributions.}
\label{fig:rens}
\end{figure}

We now compute the $H^+W^+$ and $H^+G^+$ mixing effects and 
renormalize the theory.
A set of diagrams contributes to $e^+e^- \to H^+e^-\bar\nu$
in which a $W$ boson or charged Goldstone
boson $G^+$ is radiated and turns into an $H^+$ through 
renormalized mixing diagrams.
These are shown in Figs.~\ref{fig:rent} and \ref{fig:rens} for the
$t$- and $s$-channel processes, respectively, along with the $W^+H^-V$
coupling counterterm (denoted by an $X$) which renormalizes the 
$W^+H^-V$ vertex.
We neglect all diagrams that are proportional to the electron or neutrino 
mass.

\begin{figure}
\resizebox{8.5cm}{!}{\includegraphics*[55,630][330,750]{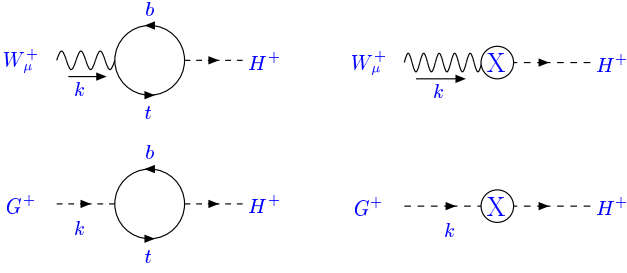}}
\caption{Top and bottom quark contributions to the $W^+H^+$ and 
$G^+H^+$ mixing (left) and the corresponding counterterms (right).}
\label{fig:2point}
\end{figure}

The top and bottom quark loops that give rise to the $W^+H^+$ and
$G^+H^+$ mixing are shown in Fig.~\ref{fig:2point}, along with their
counterterms.
The $W^+H^+$ mixing diagram is denoted by
$-i k^{\mu} \Sigma_{W^+H^+}(k^2)$, where $k$ is the incoming momentum of
the
$W^+$ boson and $H^+$ is outgoing.  The Hermitian conjugate diagram, with 
$H^-$ outgoing, has the opposite sign.
The $G^+H^+$ mixing diagram is denoted by 
$+i \Sigma_{G^+H^+}(k^2)$, where $k$ is the incoming momentum of the
$G^+$ and $H^+$ is outgoing.  The Hermitian conjugate diagram,
with $H^-$ outgoing, is the same.  The renormalized mixing two-point functions
are denoted with a hat.

Because the process $e^+e^- \to H^+ e^- \bar \nu$ is zero at tree level,
the renormalization procedure is quite simple.  
In our discussion below, we follow the same on-shell renormalization 
scheme used in Ref.~\cite{Arhrib}.  In fact, we need impose
only the following two renormalization conditions.  First, the renormalized
tadpoles are set to zero.  
Second, the real part of the renormalized $W^+H^+$ mixing is set to zero
when $H^+$ is on mass shell:
        \begin{equation}
        {\rm Re} \, \hat \Sigma_{W^+H^+} (m^2_{H^{\pm}}) = 0.
        \label{eq:rencond}
        \end{equation}
This fixes the counterterm for $W^+H^+$ mixing shown in Fig.~\ref{fig:2point}:
        \begin{equation}
        m_W \sin\beta \cos\beta \, \delta c 
        = {\rm Re} \, \Sigma_{W^+H^+} (m^2_{H^{\pm}}),
        \label{eq:deltac0}
        \end{equation}
where $\delta c$ is a combination of counterterms (see, e.g.,
Ref.~\cite{Dabelstein} for the definitions of the counterterms):
        \begin{equation}
        \delta c \equiv
        \delta Z_{H_1} - \delta Z_{H_2} - \delta v_1/v_1
        + \delta v_2 /v_2.
        \label{eq:deltac}
        \end{equation}
The renormalized $G^+H^+$ mixing two-point function, 
$\hat \Sigma_{G^+H^+}(k^2)$, is fixed in terms of the
renormalized $W^+H^+$ mixing two-point function 
$\hat \Sigma_{W^+H^+}(k^2)$ by the Slavnov-Taylor identity
(see, e.g., Refs.~\cite{SlavnovTaylor,SlavnovTaylor2} for details):
        \begin{equation}
        k^2 \hat \Sigma_{W^+H^+}(k^2) - m_W \hat \Sigma_{G^+H^+}(k^2) = 0.
        \label{eq:ST}
        \end{equation}
Thus Eq.~(\ref{eq:rencond}) also fixes the counterterm for $G^+H^+$
mixing shown in Fig.~\ref{fig:2point}.

There is also a counterterm for the $W^{+\mu}V^{\nu}H^-$
vertex (all particles incoming), given by
$-i g_V^G m_W \sin\beta \cos\beta \, \delta c \, g^{\mu\nu}$
and denoted by the $X$ in the last diagram of 
Figs.~\ref{fig:rent} and \ref{fig:rens}.
The coupling $g_V^G$ is defined in Appendix~\ref{sec:notation}, 
Eq.~(\ref{eq:gcouplings}).  
The counterterm for the Hermitian conjugate vertices, with 
$W^{-\mu}V^{\nu}H^+$ incoming, is identical.  This counterterm 
$\delta c$ is also fixed by the condition in Eq.~(\ref{eq:rencond}).

Applying these renormalization conditions, we find that the sum of the
diagrams in Figs.~\ref{fig:rent} and \ref{fig:rens} reduces to
a quite simple result for both the $t$-channel and $s$-channel processes.
See Appendix~\ref{sec:sum} for a detailed derivation. 
For the $t$-channel process, the sum of the diagrams in Fig.~\ref{fig:rent}
is:
        \begin{eqnarray}
        i \mathcal{M}_{\rm ren}^t &=&
        i e^2 g_W 
        \left[- g_V^G \Sigma_{W^+H^+}(m^2_{H^{\pm}}) g^{\mu\nu}\right]
        \label{eq:sumt}
        \\ && \hskip-1cm \times
        \frac{\bar v(e^+_i) \gamma_{\mu} P_L v(\bar\nu_f)
        \bar u(e^-_f) \gamma_{\nu}(g^{eL}_V P_L + g^{eR}_V P_R) u(e^-_i)}
        {[k_1^{t2} - m_W^2][k_2^{t2} - m_V^2]}.
        \nonumber
        \end{eqnarray}
For the $s$-channel process, the sum of the diagrams in Fig.~\ref{fig:rens} is:
        \begin{eqnarray}
        i \mathcal{M}_{\rm ren}^s &=&
        i e^2 g_W
        \left[- g_V^G \Sigma_{W^+H^+}(m^2_{H^{\pm}}) g^{\mu\nu}\right]
        \label{eq:sums}
        \\ && \hskip-1cm \times
        \frac{\bar u(e^-_f) \gamma_{\mu} P_L v(\bar\nu_f)
        \bar v(e^+_i) \gamma_{\nu} (g_V^{eL} P_L + g_V^{eR} P_R) u(e^-_i)}
        {[k_1^{s2} - m_W^2 + im_W\Gamma_W][k_2^{s2} - m_V^2]}.
        \nonumber
        \end{eqnarray}
Comparing these expressions to 
Eqs.~(\ref{eq:Mformfactors_t}) and (\ref{eq:Mformfactors_s}), the sum of the 
counterterm and wavefunction renormalization diagrams can be written as
a contribution to the form factor $G_V$, for both the $s$- and $t$-channel
processes:
        \begin{equation}
        G_V^{\rm tot} = G_V^{\rm loop} - g_V^G \Sigma_{W^+H^+}(m^2_{H^{\pm}}).
        \label{eq:sum}
        \end{equation}
The explicit expression for $\Sigma_{W^+H^+}(k^2)$ is given in 
Appendix~\ref{sec:matrixelements}.

One possible concern is that including the fixed $W$ decay width 
in the $W$ propagators for the $s$-channel diagrams might
spoil the gauge invariance of the calculation.
In our calculation, we use the 
``factorization scheme'', which is guaranteed to be gauge independent.
Following, e.g., the discussion in Ref.~\cite{nnHrcs},
the one-loop matrix element in the factorization scheme is given 
by\footnote{For processes that are nonzero at tree level, care must be
taken to avoid double-counting the $W$ width.  This is not a concern here
since the tree level matrix element is zero.}
\begin{equation}
        \mathcal{M} = \frac{k_1^{s2} - m_W^2}{k_1^{s2} - 
                        m_W^2 + i m_W \Gamma_W}
                \mathcal{M}_{\Gamma_W = 0},
\end{equation}
which is exactly the relation that we used to obtain
Eqs.~(\ref{eq:Mformfactors_s}) and (\ref{eq:sums}).

\subsection{Polarized cross sections}

To compute the polarized cross sections, we first define the
following combinations of form-factors: 
        \begin{equation}
        (G,H,F)^{t,s}_{L,R}=\sum_{V=\gamma,Z} 
        \frac{e^2 g_W g_{V}^{eL,R} (G,H,F)_V}
        {[(k_1^{t,s})^2-m_W^2][(k_2^{t,s})^2-m_V^2]}.
        \label{eq:LRformfactors}
        \end{equation}
The square of the matrix element is then given as follows.  
Because the $W$-boson couples only to left-handed fermions, 
the $t$-channel diagrams contribute only to $\mathcal{M}(e^+_Re^-_{L,R})$.
Similarly, because of the vector coupling of $V=\gamma,Z$, 
the $s$-channel diagrams contribute only to 
$\mathcal{M}(e^+_Re^-_L)$ and $\mathcal{M}(e^+_Le^-_R)$.
Thus, only the square of $\mathcal{M}(e^+_Re^-_L)$ will 
involve interference between the $s$- and $t$-channel diagrams, 
and $\mathcal{M}(e^+_Le^-_L)$ is zero.
In particular, we have:
\begin{eqnarray}
        &&|{\cal{M}}(e^+_{R} e^-_{L})|^2=K_{L}^t+ K_{L}^s +K^{st},
        \quad
        |{\cal{M}}(e^+_{L} e^-_{R})|^2=K_{R}^s,
        \nonumber \\
        &&|{\cal{M}}(e^+_{R} e^-_{R})|^2=K_{R}^t,
        \quad
        |{\cal{M}}(e^+_L e^-_L)|^2=0,
\end{eqnarray}
where $K_{L,R}^t$, $K_{L,R}^s$ and $K^{st}$
are given in Appendix~\ref{sec:matrixelements} in 
terms of the form-factors in Eq.~(\ref{eq:LRformfactors}).

\section{\label{sec:numerics}Numerical results}
In this section, we present our numerical results for the cross 
section of $e^+e^-\rightarrow H^+e^-\bar{\nu}$.  
We have used the LoopTools package \cite{LoopTools} to compute
the one-loop integrals.
The electron and positron beams are assumed to be unpolarized
unless specified explicitly.
Also, we plot the cross section for $e^+e^- \to H^+e^-\bar\nu$ only.  The 
cross section for the charge-conjugate process is the same
(for unpolarized beams); adding 
them together doubles the cross sections shown.

\begin{figure}
\resizebox{8.5cm}{!}{\rotatebox{270}{\includegraphics{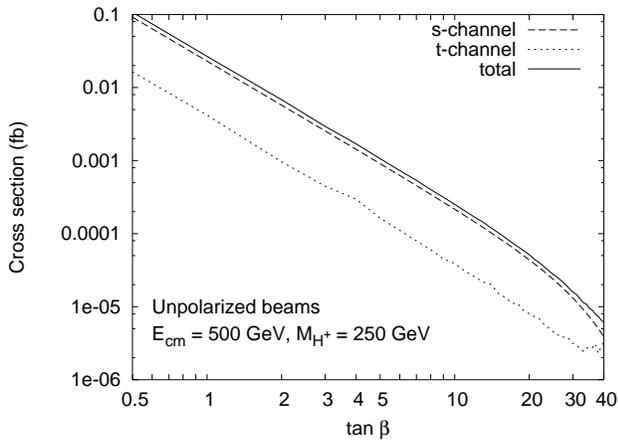}}}
\caption{Cross section for $e^+e^- \to H^+e^-\bar\nu$ as a function of 
$\tan\beta$ for $\sqrt{s} = 500$ GeV and $m_{H^{\pm}} = 250$ GeV.
We show the total cross section (solid line), $s$-channel 
contribution (dashed line) and $t$-channel contribution (dotted line).}
\label{fig:500tanb}
\end{figure}

\begin{figure}
\resizebox{8.5cm}{!}{\rotatebox{270}{\includegraphics{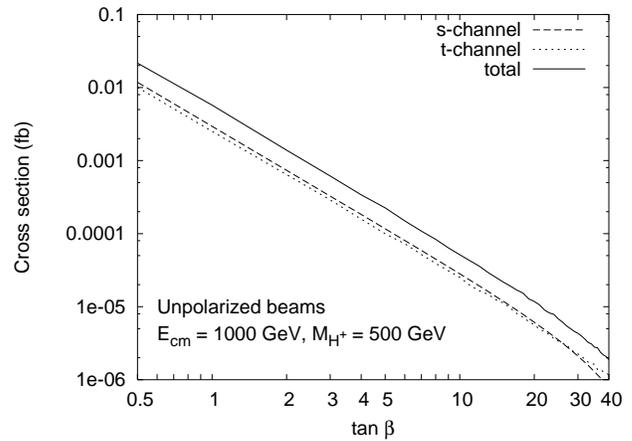}}}
\caption{As in Fig.~\ref{fig:500tanb} but for $\sqrt{s} = 1000$ GeV
and $m_{H^{\pm}} = 500$ GeV.}
\label{fig:1000tanb}
\end{figure}

We first show the $\tan\beta$ dependence of the cross section for 
fixed $m_{H^\pm}=\sqrt{s}/2$ in Figs.~\ref{fig:500tanb} ($\sqrt{s}=500$ GeV)
and \ref{fig:1000tanb} ($\sqrt{s}=1000$ GeV).
The cross section falls like $(\tan\beta)^{-2}$ due to the factor
of $y^L_{Htb} \propto \cot\beta$ in the matrix elements.  
At very large $\tan\beta$ values the cross 
section begins to turn upward again due to terms proportional to
$y^R_{Htb} \propto \tan\beta$ in the matrix elements.  This effect is barely 
visible at $\tan\beta = 40$ in Fig.~\ref{fig:500tanb}.
At a 500 GeV LC, the cross section for a charged Higgs boson with 250 GeV mass 
is larger than 0.01 fb (corresponding to 10 events for integrated luminosity
${\cal L}=500\ {\rm fb}^{-1}$ when both $H^+$ and $H^-$ 
are taken into account) 
only for small $\tan\beta < 1.7$. 

\begin{figure}
\resizebox{8.5cm}{!}{\rotatebox{270}{\includegraphics{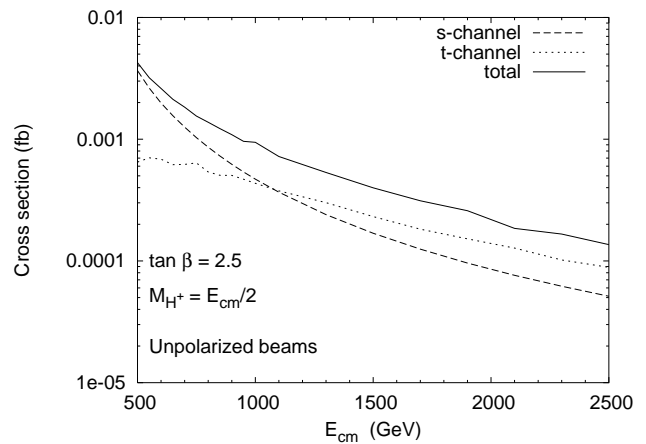}}}
\caption{Cross section as a function of $\sqrt{s}$ for 
$m_{H^{\pm}} = \sqrt{s}/2$ and $\tan\beta = 2.5$, to show different 
behavior of $s$-channel and $t$-channel contributions.}
\label{fig:sqrts}
\end{figure}

From Figs.~\ref{fig:500tanb} and \ref{fig:1000tanb} 
we can also see that at $\sqrt{s}=500$ GeV, the $s$-channel contribution 
dominates the total cross section, while at $\sqrt{s}=1000$ GeV, the $s$- 
and $t$- channel contributions become comparable.  This center-of-mass energy
dependence of the cross section 
can be seen in Fig.~\ref{fig:sqrts}, which shows the 
$s$-channel (dashed line) and $t$-channel (dotted line)
contributions versus center-of-mass energy for 
$m_{H^{\pm}}=\sqrt{s}/2$ and $\tan\beta=2.5$.  
The $t$-channel contribution dominates when $\sqrt{s}>1050$ GeV.
This crossover happens at a higher center-of-mass energy
than in the case of the similar process 
$e^+e^-\rightarrow \nu \bar{\nu}A^0$, for which the 
$t$-channel dominates when $\sqrt{s}>670$ GeV \cite{nunuA}.  In general,
the $t$-channel contribution becomes more important than the $s$-channel
contribution at larger $\sqrt{s}$ values because of the 
propagator suppression $\sim 1/s$ of the $s$-channel contribution.
In the process 
$e^+e^-\rightarrow \nu \bar{\nu}A^0$, there is an additional enhancement 
of the $t$-channel contribution from the 
fact that the $W^+ e^- \bar{\nu}$ coupling
in the $t$-channel process is larger than the $Z e^+e^-$ and
$Z \nu \bar{\nu}$ couplings in the $s$-channel process.  
This makes the $t$-channel contribution start to  
dominate at a smaller $\sqrt{s}$ value in $e^+e^-\rightarrow \nu \bar{\nu}A^0$
than in $e^+e^-\rightarrow H^+ e^- \bar{\nu}$, where such 
additional enhancement from the relevant couplings does not occur.

\begin{figure}
\resizebox{8.5cm}{!}{\rotatebox{270}{\includegraphics{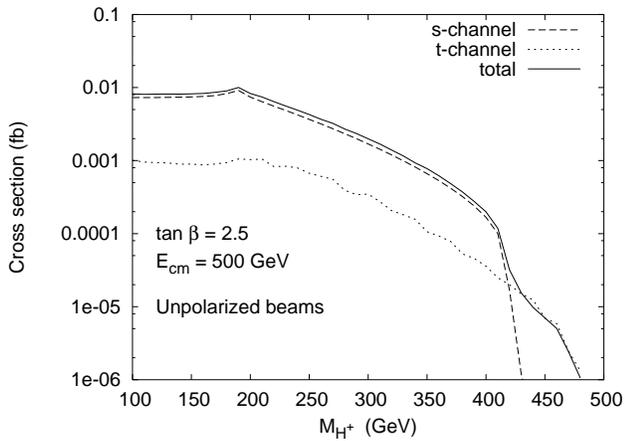}}}
\caption{Cross section as a function of $m_{H^{\pm}}$ for $\tan\beta = 2.5$
and $\sqrt{s} = 500$ GeV.}
\label{fig:2.5mhpm}
\end{figure}

\begin{figure}
\resizebox{8.5cm}{!}{\rotatebox{270}{\includegraphics{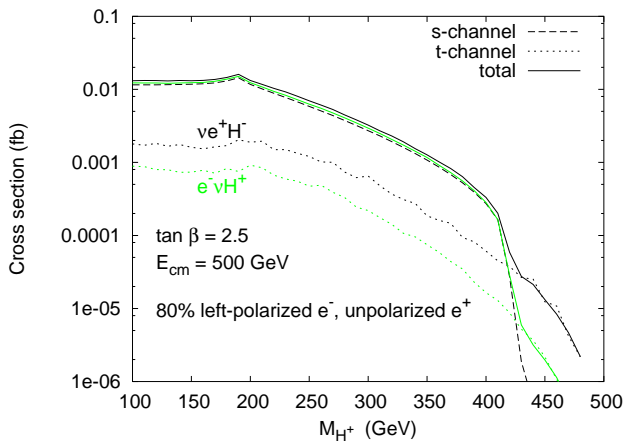}}}
\caption{As in Fig.~\ref{fig:2.5mhpm} but with an 80\% left-polarized 
$e^-$ beam.  The $t$-channel contribution is now different for 
the $H^+e^-\bar\nu$ and $H^-e^+\nu$ final states, as indicated by the
green (gray) and black lines, respectively.  Notice that there
are two solid lines for total cross section, corresponding to 
$H^+e^-\bar\nu$ and $H^-e^+\nu$ respectively. 
}
\label{fig:2.5mhpmLH}
\end{figure}

We also show the charged Higgs mass dependence of the cross section for an
unpolarized electron beam (Fig.~\ref{fig:2.5mhpm}) and an 80\% left-polarized 
electron beam (Fig.~\ref{fig:2.5mhpmLH}).  
With a polarized electron beam (and unpolarized positron beam), the 
$t$-channel contribution to the $e^+e^- \to H^+e^-\bar\nu$ cross section
is different from that to the $e^+e^- \to H^-e^+\nu $ cross section.
This is because for $e^+e^- \to H^+e^-\bar\nu$, the $t$-channel
$W$ boson couples to the positron beam, while for 
$e^+e^- \to H^-e^+\nu$ it couples to the electron beam.
Left-polarizing the electron beam thus has a much more sizable effect 
on the $t$-channel $e^+e^- \to H^-e^+\nu$ cross section, enhancing
it significantly.  Left-polarizing the electron beam causes a slight
suppression of the $t$-channel $e^+e^- \to H^+e^-\bar\nu$ cross section,
due to the relative size of the sum of the photon and $Z$ exchange 
diagrams for left- and right-handed electrons.
The $s$-channel contribution is identical for the two final states;
it is enhanced by about 50\% with an 80\% left-polarized $e^-$
beam, which is consistent with the results for the process
$e^+e^-\rightarrow H^+W^-$ \cite{HW}.

\begin{figure}
\resizebox{8.5cm}{!}{\rotatebox{270}{\includegraphics{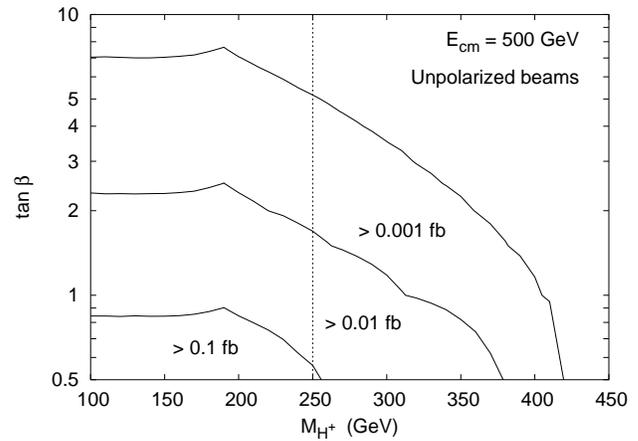}}}
\caption{Cross section in the $m_{H^{\pm}}-\tan\beta$ plane for 
unpolarized beams with $\sqrt{s} = 500$ GeV.  The vertical dotted line
indicates the pair production threshold, 
$m_{H^{\pm}} = \sqrt{s}/2 = 250$ GeV.}
\label{fig:contour}
\end{figure}

We summarize the results of Figs.~\ref{fig:500tanb} and \ref{fig:2.5mhpm}
in Fig.~\ref{fig:contour}, where we show contours of cross section for
$e^+e^- \to H^+e^-\bar\nu$ in the $m_{H^{\pm}}-\tan\beta$ plane.
A cross section above 0.1 fb corresponds to 100 charged Higgs events
(combining $H^+$ and $H^-$) in an integrated luminosity of 500 fb$^{-1}$.
This occurs mainly only for $m_{H^{\pm}} < \sqrt{s}/2$, and thus would
not aid in the charged Higgs discovery.  It could, however, be useful for
measuring $\tan\beta$ at low $\tan\beta$ values, due to the strong
$(\tan\beta)^{-2}$ dependence of the cross section.  The
$e^+e^- \to H^+e^-\bar\nu$ events could be separated from 
$e^+e^- \to H^+H^-$ events by using the fact that $H^- \to e^-\bar\nu$
is suppressed by the tiny electron Yukawa coupling.  A cross section
above 0.01 fb corresponds to 10 charged Higgs events (again combining
$H^+$ and $H^-$) in 500 fb$^{-1}$, and is probably the limit of relevance
of this process.  It offers some reach for $m_{H^{\pm}} > \sqrt{s}/2$
for low $\tan\beta$ values below 1.7.  The statistics could be 
roughly doubled by including the $H^+\mu^-\bar\nu$ and $H^-\nu\mu^+$
final states, which receive only $s$-channel contributions and have 
the same $s$-channel cross section as the process with an electron
or positron in the final state.

\begin{figure}
\resizebox{8.5cm}{!}{\rotatebox{270}
{\includegraphics{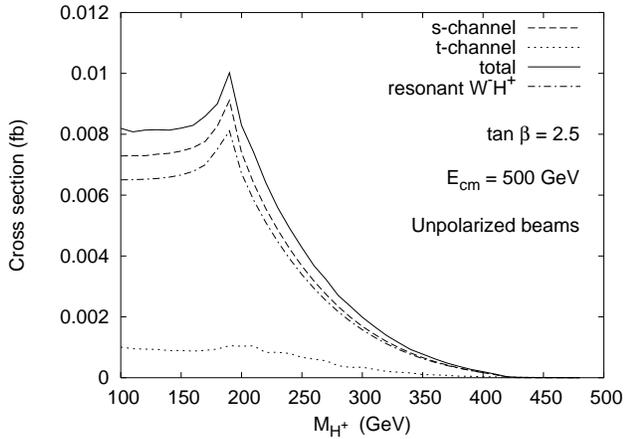}}}
\caption{Comparison of the cross section for $e^+e^- \to H^+e^-\bar\nu$
with the resonant process $e^+e^- \to H^+W^-$, with $W^- \to e^- \bar \nu$.  
Here we plot the cross section on a linear scale so that the difference
between the full $s$-channel contribution and the resonant sub-diagrams
will be visible.}
\label{fig:HWcompare}
\end{figure}

Since the $s$-channel contribution dominates the total cross 
section for $\sqrt{s} \lsim 1000$ GeV,
we show the comparison of the cross section for  $e^+e^- \to H^+e^-\bar\nu$
with that of the resonant process $e^+e^- \to H^+W^-$ 
(with $W^- \to e^- \bar \nu$) in Fig.~\ref{fig:HWcompare}.
In the $e^+e^- \to H^+W^-$ cross section calculation we include only the top 
and bottom quark loops and we use the tree-level result for
the branching ratio of $W^- \to e^- \bar \nu$ in order
to make a consistent comparison.
The resonant contribution is about 80\% of the full $s$-channel cross
section.

\section{\label{sec:modeldependence}Model dependence}

Here we discuss the model dependence of our results, and show how they
can be extended beyond our current calculations.

\subsection{Higgs potential}

Since we have included only the top and bottom quark loops in our 
calculation, our results are valid regardless of the form of the 
Higgs potential for the 2HDM.  Once the contributions of gauge and Higgs
boson loops are included, however, the cross section will depend on 
the form of the Higgs potential through the Higgs boson self-couplings.
As shown in Ref.~\cite{Kanemura}, the effect of the gauge and Higgs boson loops
on the $e^+e^- \to H^+W^-$ cross section is negligible unless the Higgs
boson self couplings are very large; even then, the gauge and Higgs boson 
loops only become important for $\tan\beta \gsim 4$, where the cross
section has already fallen off by more than an order of magnitude compared to
$\tan\beta = 1$.  Thus, even if the Higgs boson 
self-couplings are large, we do not expect the gauge and Higgs boson loops 
to be important when the cross section is large enough to be observed.

\subsection{Type I vs.\ Type II 2HDM}
  
In our numerical results we have assumed that the Higgs sector is
the Type II 2HDM, in which one Higgs doublet gives mass to the up-type
quarks and the other Higgs doublet gives mass to the down-type quarks
and the charged leptons.
In the Type I 2HDM, one Higgs doublet gives mass
to all fermions.  In this case, all our results remain the same except that
$y^R_{Htb}$ in Eq.~(\ref{eq:yLyR}) is modified to
\begin{equation}
        y^R_{Htb} = \frac{m_b \cot\beta}{\sqrt{2} m_W s_W}
        \qquad {\rm (Type \ I)}.
        \label{eq:yLyRTypeI}
\end{equation}
The effect of this is that the curves in Fig.~\ref{fig:500tanb} do not
start to flatten out at high $\tan\beta$.  The results at low $\tan\beta$
are unchanged.  Thus, to a very good approximation, our results are valid
also for the Type I 2HDM, except at large $\tan\beta$.

\subsection{Higgs sector extensions beyond the 2HDM}

We may also consider the effects of extending the Higgs sector beyond
the 2HDM.  Adding neutral Higgs singlets can only affect the gauge and Higgs
boson loops, which we have argued are unlikely to be experimentally 
relevant.  Adding charged Higgs bosons via additional doublets or charged
singlets leads to a Higgs sector with more than one charged Higgs boson,
the lightest of which will in general be a mixture of the various gauge
eigenstates.  To avoid Higgs-mediated flavor changing neutral currents,
fermions of each charge should couple to no more than one Higgs doublet
\cite{noTypeIII}.
Thus the mixing will in general lead to a suppression of the lightest 
$H^{\pm}$ couplings to top and bottom quarks; the effect will be 
the same as increasing $\tan\beta$ in the Type I model.

Adding larger Higgs multiplets leads to more interesting effects.
For example, adding a Higgs triplet with a nonzero vacuum expectation
value leads to a nonzero $W^+Z H^-$ coupling \emph{at tree level}.
This tree-level $W^+ZH^-$ coupling is proportional to the triplet
vacuum expectation value $v_3$:\footnote{The vacuum expectation value
is normalized according to $\langle \phi^0 \rangle = v_3/\sqrt{2}$ for
a complex triplet and $\langle \phi^0 \rangle = v_3$ for a real triplet.}
\begin{equation}
  g_{W^+ZH^-} = 
\left\{ 
\begin{array}{lcl}
      g^2 (c^2_W-2) v_3 / \sqrt{2} c_W
    & \qquad & {\rm complex \ triplet} \\
      g^2 c_W v_3
    & \qquad & {\rm real \ triplet}.
    \end{array}
\right.
\end{equation}
The triplet vacuum expectation value
is forced to be small by the experimental
constraints on weak isospin violation:
\begin{equation}
  \delta \rho = \left\{ \begin{array}{lcl}
      - 2 v_3^2 / v_{SM}^2 & \qquad & {\rm complex\ triplet} \\
        4 v_3^2 / v_{SM}^2 & \qquad & {\rm real\ triplet}.
\end{array}
\right.
\end{equation}
The current constraint
on contributions to the $\rho$ parameter from new physics is
$-0.0016 < \Delta \rho < 0.0058$ at the $2\sigma$ level \cite{PDG}.  This
leads to an upper bound
on the \emph{tree-level} cross section for $e^+e^- \to H^+e^-\bar\nu$
of 0.0035 fb for a real triplet, or 0.0026 fb for a complex triplet, with
$m_{H^+} = 250$ GeV at $\sqrt{s} = 500$ GeV and assuming unpolarized
beams.
This corresponds to 3.5 events (counting $H^+$ and $H^-$ final states)
in 500 fb$^{-1}$ of integrated
luminosity for a real triplet, and 2.6 events for a complex
triplet, both of which are too small to be observable.
From Fig.~\ref{fig:500tanb}, these cross sections become comparable to 
the loop-induced cross section in the 2HDM at $\tan\beta \simeq 3-3.5$.

\subsection{MSSM}

In the MSSM, this process will get contributions from SUSY loops, just
as $e^+e^- \to H^+W^-$ does.  In the latter case, the SUSY loops could
increase the cross section by 50-100\% at low $\tan\beta$
if the SUSY particles are relatively light \cite{HWupdate}.
Since the resonant $e^+e^- \to H^+W^-$ subprocess dominates the cross section
at $\sqrt{s} \lsim 1000$ GeV, we expect this enhancement to carry over.
In the MSSM, however, $\tan\beta$ is limited to be above 2.4 \cite{LEP2},
which already leads to a somewhat smaller cross section.

\begin{figure}
\resizebox{8.5cm}{!}{\rotatebox{270}{\includegraphics{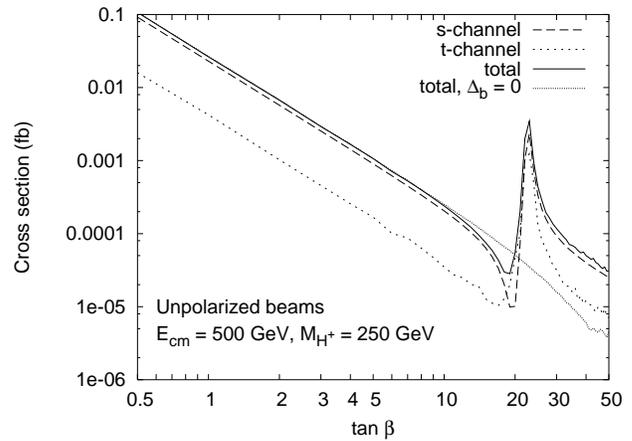}}}
\caption{Cross section for $e^+e^- \to H^+e^-\bar\nu$ as a function of 
$\tan\beta$ for
$\sqrt{s} = 500$ GeV and $m_{H^{\pm}} = 250$ GeV.  
The MSSM parameters are specified in the text.
In the region near $\Delta_b=-1$, we cut off the divergence by
requiring $e |y^R_{Htb}| \leq 4\pi$.
The dotted line shows the total cross section for $\Delta_b = 0$.}
\label{fig:Deltab}
\end{figure}

A different source of SUSY corrections to the $e^+e^- \to H^+e^-\bar \nu$
cross section is the SUSY radiative correction to the bottom quark
Yukawa coupling, parameterized by $\Delta_b$ \cite{Deltab}.
This correction arises from a coupling of the second Higgs doublet $\Phi_2$
to the bottom quark induced at one-loop by SUSY-breaking terms:
\begin{equation}
        -\mathcal{L}_{\rm Yukawa} \simeq h_b \Phi_1^0 \bar b b
        + (\Delta h_b) \Phi_2^0 \bar b b.
        \label{eq:LYuk}
\end{equation}
While $\Delta h_b$ is one-loop suppressed compared to $h_b$, 
for sufficiently large $\tan\beta \equiv v_2/v_1$ the contribution
of both terms in Eq.~(\ref{eq:LYuk}) to the $b$ quark mass can be
comparable in size.  This leads to a large modification of the tree-level
relation for the bottom quark mass,
\begin{equation}
        m_b = \frac{h_b v_1}{\sqrt{2}}(1 + \Delta_b),
\end{equation}
where $\Delta_b \equiv (\Delta h_b) \tan\beta/h_b$.  The correction
$\Delta_b$ comes from two main sources: (1) a bottom squark--gluino loop,
which depends on the masses $m_{\tilde b_{1,2}}$ of
the two bottom squarks and the gluino mass $m_{\tilde g}$; and
(2) a top squark--Higgsino loop, which depends on the masses 
$m_{\tilde t_{1,2}}$ of the two top squarks and the Higgsino mass parameter
$\mu$.  Neglecting contributions proportional to the electroweak gauge
couplings, we have explicitly \cite{Deltab}:
\begin{eqnarray}
        \Delta_b &\simeq& \frac{2 \alpha_s}{3\pi} m_{\tilde g} \mu \tan\beta
        \ I(m_{\tilde b_1},m_{\tilde b_2},m_{\tilde g})
        \nonumber \\
        &&+ \frac{Y_t}{4\pi}A_t \mu \tan\beta
        \ I(m_{\tilde t_1},m_{\tilde t_2},\mu).
        \label{eq:Deltab}
\end{eqnarray}
Here $A_t$ is the trilinear coupling in the top squark sector, 
$\alpha_s = g_s^2/4\pi$, and $Y_t \equiv h_t^2/4\pi$.
The loop function $I$ is positive definite.  Since the Higgs coupling
$\Delta h_b$ is a manifestation of SUSY breaking, it does not decouple
in the limit of large SUSY breaking mass parameters.  In fact, if all
SUSY breaking mass parameters and $\mu$ are scaled by a common factor,
$\Delta_b$ remains constant.
These corrections modify $y^R_{Htb}$ in Eq.~(\ref{eq:yLyR}):
\begin{equation}
        y^R_{Htb} = \frac{m_b \cot\beta}{\sqrt{2}m_Ws_W}\frac{1}{1+\Delta_b}
        \qquad ({\rm MSSM}).
\end{equation}
To explore the impact of these effects, we plot
the cross section for $e^+e^- \to H^+e^-\bar\nu$ in Fig.~\ref{fig:Deltab}
with the specific choices
of $m_{H^+}=250$ GeV, $\sqrt{s}=500$ GeV,
$\mu=-2$ TeV, $m_{\tilde{b}_R}=525$ GeV, 
$m_{\tilde{g}}=
m_{\tilde{b}_L}=m_{\tilde{t}_{L,R}}=1$ TeV, 
and $A_t=A_b=\mu/\tan\beta+\sqrt{6} m_{\tilde{t}_L}$ (corresponding to 
maximal-mixing in the top squark sector),  and compare
to the result that would be obtained without including $\Delta_b$.
For these input parameters, $\Delta_b$ varies 
between $-\tan\beta/26$ and $-\tan\beta/22$ for $\tan\beta$ 
between 1 and 50.       
For $\Delta_b \simeq -1$ the resulting $y^R_{Htb}$ becomes quite
large, making our perturbative results unreliable.  To control such 
effects we cut off the divergence in this region by requiring
$e|y^R_{Htb}| \leq 4\pi$.
The dip in the cross section at $\tan\beta \simeq 20$ is due to 
destructive interference between various terms,
which changes to constructive interference
for $\tan\beta > 22$ where $\Delta_b < -1$ and $y^R_{Htb}$ has the
opposite sign than in the SM case.
While the modification of the cross section due to $\Delta_b$ can
be quite significant, it occurs at large $\tan\beta$ where the cross
section is already very small, and thus is unlikely to be observable 
for perturbative values of $y^R_{Htb}$.  

\section{\label{sec:conclusions}Conclusions}

We have computed the cross section for $e^+e^- \to H^+e^-\bar\nu$,
from one-loop diagrams involving top and bottom quarks in the Type II
two-Higgs-doublet model.  
This process is interesting because it offers an opportunity to
produce the charged Higgs boson in $e^+e^-$ collisions for charged Higgs
masses above half of the collider center-of-mass energy.
Because this process first appears at the one-loop
level, however, its cross section is very small.
At low values of $\tan\beta \sim 1-2$ the cross section can reach the 
0.01 fb level at $\sqrt{s} = 500$ GeV and $m_{H^{\pm}} = 250$ GeV,
leading to 10 events in 500 fb$^{-1}$ when the cross sections for 
$H^+$ and $H^-$ production are combined.  Using an 80\% left-polarized
$e^-$ beam increases the cross section by about 50\%.
The cross section falls with increasing $\tan\beta$ like $(\tan\beta)^{-2}$.  

The process $e^+e^- \to H^+e^-\bar\nu$ receives contributions
from $s$-channel and $t$-channel diagrams, which behave differently 
with increasing center-of-mass energy.  The $s$-channel contribution 
dominates at low $\sqrt{s} \lsim 1$ TeV, while the $t$-channel contribution
dominates at higher energies.  The $s$-channel contribution is in turn
dominated (at the 80\% level) by the resonant sub-process, 
$e^+e^- \to H^+W^-$, with $W^- \to e^- \bar\nu$.
Because of this, in the context of the MSSM we expect the effects of
light superpartners calculated for the $e^+e^- \to H^+W^-$ cross
section to carry over to the process $e^+e^- \to H^+e^-\bar\nu$.  Light
superpartners can lead to an increase in the $e^+e^- \to H^+W^-$
cross section by 50-100\% at low $\tan\beta$.

The final state containing $e^- \bar\nu$ is experimentally attractive 
because it is easy to tag
compared to hadronic $W$ decays.  To increase the statistics, one 
could also include the $e^+e^- \to H^+ \mu^- \bar \nu$ channel,
which comes only from the $s$-channel diagrams.
This would roughly double the cross section
at $\sqrt{s} = 500$ GeV, where the $t$-channel process does not contribute
significantly.  The processes $e^+e^- \to H^+ q \bar q^{\prime}$ 
for the first two generations of quarks 
could also be included (analogous to including hadronic $W$ decays);
it would then contribute another six 
times the $s$-channel cross section of the electron mode.  
For the third generation, $e^+e^- \to H^+ \tau^- \bar \nu$ and 
$e^+e^- \to H^+ \bar{t} b$, 
however, the tree level contributions dominate due to the 
relatively large Yukawa couplings.  Such channels
have already been studied in \cite{tbHtaunuH}.

For $m_{H^{\pm}} < \sqrt{s}/2$, where the charged Higgs boson can
be pair produced, the process $e^+e^- \to H^+e^-\bar\nu$ could still
be useful to measure $\tan\beta$ due to the strong $(\tan\beta)^{-2}$ 
dependence of the cross section.  This process should be separable
from the pair production process since the $H^-$ decay to $e^- \bar \nu$
is suppressed by the tiny electron Yukawa coupling.

\begin{acknowledgments}
We are grateful to 
Sven Heinemeyer for helpful comments.
H.E.L. and S.S. thank the Aspen Center for Physics where part of this
work was finished.
T.F. is supported by U.S. Department of Energy grant
No. DE-FG03-91ER40674 and by the Davis Institute for High
Energy Physics.
H.E.L. is supported in part by the U.S.~Department of Energy
under grant DE-FG02-95ER40896
and in part by the Wisconsin Alumni Research Foundation.
S.S. is supported by the DOE under grant DE-FG03-92-ER-40701 and 
by the John A. McCone Fellowship.
\end{acknowledgments}

\appendix

\section{\label{sec:notation}Notation and conventions}

For the one-loop integrals we
follow the notation of Ref.~\cite{LoopTools}.
The one-point integral is:
\begin{equation}
        \frac{i}{16 \pi^2}A(m^2) = \int \! \frac{d^Dq}{(2\pi)^D}
        \frac{1}{(q^2 - m^2)},
\end{equation}
where $D$ is the number of dimensions.
The two-point integrals are:
\begin{eqnarray}
        &&\frac{i}{16\pi^2}\left\{B_0, k^{\mu}B_1 \right\}(k^2,m_0^2,m_1^2)
        = \\
        &&\hskip1cm \int \! \frac{d^Dq}{(2\pi)^D}
        \frac{ \{1, q^{\mu} \} }{(q^2 - m_0^2)((q+k)^2 - m_1^2)}.
        \nonumber
\end{eqnarray}
The three-point integrals are:
\begin{eqnarray}
        &&\frac{i}{16 \pi^2}\left\{ C_0, C^{\mu}, C^{\mu\nu} \right\}
        = \\
        &&\hskip0cm \int \! \frac{d^Dq}{(2\pi)^D}
        \frac{ \{1, q^{\mu}, q^{\mu}q^{\nu} \} }
        {(q^2 - m_0^2)((q+k_1)^2 - m_1^2)((q+k_2)^2 - m_2^2)},
        \nonumber
\end{eqnarray}
where the tensor integrals are decomposed in terms of scalar components
as
\begin{eqnarray}
        C^{\mu} &=& k_1^{\mu}C_1 + k_2^{\mu}C_2  \nonumber \\
        C^{\mu\nu} &=& g^{\mu\nu} C_{00} + k_1^{\mu}k_1^{\nu}C_{11}
        + k_2^{\mu}k_2^{\nu}C_{22} \nonumber \\
        && + (k_1^{\mu}k_2^{\nu} + k_2^{\mu}k_1^{\nu})C_{12}.
\end{eqnarray}
The arguments of the scalar three-point integrals are 
$(k_1^2,(k_2-k_1)^2,k_2^2,m_0^2,m_1^2,m_2^2)$.

For couplings and Feynman rules we follow the conventions of
Ref.~\cite{HaberKane}.
The photon and $Z$ boson coupling coefficients to fermions are:
        \begin{eqnarray}
        &&g_{\gamma}^{fL}=g_{\gamma}^{fR}=-e_f,
        \\
        &&g_{Z}^{fL}=(-T_3+e_f s_W^2)/s_Wc_W, \quad
        g_{Z}^{fR}=(e_f s_W^2)/s_Wc_W,  \nonumber
        \end{eqnarray}
where the electric charges are $e_{\nu} = 0$, $e_e = -1$, $e_u = 2/3$,
and $e_d = -1/3$ and 
where $T_3=1/2$ for $\nu$, $u$ and $T_3=-1/2$ for $e$, $d$.

For the $W$ and $Z$ boson and photon couplings to fermions 
and the Goldstone boson we define: 
\begin{eqnarray}
        && g_W=-1/\sqrt{2} s_W, \quad g_{\gamma} = -e, \quad 
        g_Z = - e c_W / s_W,
        \nonumber \\
        &&g^G_{\gamma} = e, \quad g^G_Z = - e s_W / c_W.
        \label{eq:gcouplings}
\end{eqnarray}

Finally, the $H^{\pm}$ coupling coefficients to top 
and bottom quarks are:
\begin{equation}
        y^L_{Htb} = \frac{m_t \cot\beta}{\sqrt{2} m_W s_W}, \qquad
        y^R_{Htb} = \frac{m_b \tan\beta}{\sqrt{2} m_W s_W}.
        \label{eq:yLyR}
\end{equation}

\section{\label{sec:matrixelements}Matrix elements}

\subsection{Top and bottom quark contributions}

The $W^+H^+$ mixing diagram (Fig.~\ref{fig:2point})
was given in Refs.~\cite{Arhrib,Kanemura,HW}:
        \begin{eqnarray}
        \Sigma_{W^+H^+}(k^2) &=& 
        \frac{N_c \alpha}{2 \pi} g_W 
        \left[ (m_t y^L_{Htb} + m_b y^R_{Htb}) B_1 
        \right. \nonumber \\ && \qquad \left.
        + m_b y^R_{Htb} B_0 \right],
        \end{eqnarray}
with the arguments of the two-point integrals given by
$B(k^2, m_b^2, m_t^2)$.

The top and bottom quark contributions to the effective $W^+_{\mu} V_{\nu} H^-$
coupling are shown in Fig.~\ref{fig:tb}.  
The corresponding form-factors $G_V$, $H_V$ and $F_V$
can be read off from the $s$-channel 
matrix element for $e^+e^- \to W^+H^-$
given in Refs.~\cite{Arhrib,Kanemura,HW} using the following relation:
\begin{widetext}
\begin{eqnarray}
        i \mathcal{M} &=& \frac{ie}{s-m_V^2} \left\{
        G_V [g_V^{eR} \mathcal{A}_1 + g_V^{eL} \mathcal{A}_2]
        - H_V [g_V^{eR} (\mathcal{A}_3 + \mathcal{A}_5)
        + g_V^{eL} (\mathcal{A}_4 + \mathcal{A}_6)]
        + F_V [g_V^{eR} \mathcal{A}_7 + g_V^{eL} \mathcal{A}_8] \right\},
\end{eqnarray}
where the $\mathcal{A}_i$ are matrix elements defined in
Refs.~\cite{Arhrib,HW}.
The loop involving $t,b,b$ gives:
        \begin{eqnarray}
        G_V &=& \frac{e \alpha N_c g_W}{2 \pi} \left[
        -(m_t\gvdl\yl+m_b(\gvdl-\gvdr)\yr)B_0
        +2\gvdl(m_t\yl+m_b\yr)C_{00}\right. \nonumber \\
        &&\left.
        +(-\gvdl(m_t\yl+m_b\yr) k_1^2
        +(m_t\gvdl\yl-m_b\gvdr\yr)(-k_1^2 - k_1\cdot k_2))C_1
        \right.\nonumber \\
        &&\left.
        +(\gvdl(m_t\yl+m_b\yr)(-k_1^2 - k_1\cdot k_2)
        - (m_t\gvdl\yl-m_b\gvdr\yr)m_{H^{\pm}}^2)C_2
        \right. \nonumber \\
        &&\left.
        +(-m_t^3\gvdl\yl-m_t^2m_b(\gvdl-\gvdr)\yr+m_tm_b^2\gvdr\yl+
        m_t\gvdl\yl(-k_1^2 - k_1\cdot k_2))C_0 \right]
        \nonumber \\
        H_V &=& - \frac{e \alpha N_c g_W}{2 \pi} \left[
        -2\gvdl(m_t\yl+m_b\yr)(C_{12}+C_{22})-\gvdl(3m_t\yl+m_b\yr)C_2
        -(m_t\gvdl\yl-m_b\gvdr\yr)C_1
        \right. \nonumber \\
        &&\left.
        -m_t\gvdl\yl{C}_0 \right]
        \nonumber \\
        F_V &=& \frac{e \alpha N_c g_W}{2 \pi} \left[
        m_t\gvdl\yl{C}_0+(m_t\gvdl\yl+m_b\gvdr\yr)C_1
        +\gvdl(m_t\yl+m_b\yr)C_2 \right]
        \end{eqnarray}
\end{widetext}
with the arguments for the integral functions as 
$B(k_2^2,m_b^2,m_b^2)$, $C(k_1^2,k_2^2,m_{H^{\pm}}^2,m_t^2,m_b^2,m_b^2)$.
The contributions of the loop involving $t,t,b$ are given by the
substitutions: 
\begin{displaymath}
        m_t \leftrightarrow m_b, \ \ \gvdl\leftrightarrow\gvul,\ \ \ 
        \gvdr\leftrightarrow\gvur, \ \ \ \yl\leftrightarrow\yr;
\end{displaymath}
and $F_V$ gets an overall minus sign.

\subsection{Square of the matrix element}

The square of the matrix element is given as follows in terms
of the form-factors $(G,H,F)_{L,R}^{t,s}$ defined in 
Eq.~(\ref{eq:LRformfactors}).

We define the following kinematic variables:
\begin{eqnarray}
&&s=2\eminusi\cdot\eplus,\quad \hat{s}=2\eminusf\cdot\neubar,\quad 
t_1=-2\eminusi\cdot\eminusf,\nonumber \\
&&t_2=-2\eplus\cdot\neubar,\quad
u_1=-2\eminusi\cdot\neubar,\quad u_2=-2\eplus\cdot\eminusf\,, \nonumber \\
&&\eps(\eminusi,\eplus,\eminusf,\neubar)=\eps^{\mu\nu\rho\sigma}
(\eminusi)_\mu\,\eplus_\nu\,(\eminusf)_\rho\neubar_\sigma\,.
\end{eqnarray}
For the antisymmetric tensor we use the convention $\epsilon^{0123}=+1$.

The squares of the matrix elements $K^t_{L,R}$, $K^s_{L,R}$,
and $K^{st}$ are given as follows.

\begin{widetext}

$K_L^t$ is given by:
\begin{eqnarray}
K_L^t&=&4|G_L^t|^2u_1u_2
\nonumber \\
&+&|H_L^t|^2(s\shat-t_1 t_2 +u_1 u_2+su_1+\shat u_2)
(s\shat-t_1 t_2 +u_1 u_2+su_2+\shat u_1) 
\nonumber \\
&+&|F_L^t|^2(-s^2\shat^2-t_1^2 t_2^2 -u_1^2 u_2^2 
+2 s \shat t_1 t_2 +2 s \shat u_1 u_2 +t_1 t_2 u_1^2 +t_1 t_2 u_2^2 )
\nonumber \\
&+&2 {\rm Re}(G_L^tH_L^{t*})(u_1^2 u_2 + u_1 u_2^2  
+ s\shat u_1 + s\shat u_2 + 2 s u_1 u_2 + 2 \shat u_1 u_2 
-t_1 t_2 u_1 -t_1 t_2 u_2 )
\nonumber \\ 
&+&8 {\rm Im}(G_L^tH_L^{t*})\eps(\eminusi,\eplus,\eminusf,\neubar)(u_1-u_2)
\nonumber \\
&+&2 {\rm Re}(G_L^tF_L^{t*})(u_1^2 u_2 -u_1 u_2^2 -s \shat u_1 + s \shat u_2 
+t_1 t_2 u_1 - t_1 t_2 u_2 )
\nonumber \\
&-&8 {\rm Im}(G_L^tF_L^{t*})\eps(\eminusi,\eplus,\eminusf,\neubar)(u_1+u_2)
\nonumber \\
&-& {\rm Re}(H_L^tF_L^{t*})
(u_1-u_2)(s^2 \shat + s\shat^2 -u_1^2 u_2 - u_1 u_2^2 
+ s \shat u_1 + s \shat u_2 
\nonumber \\
&&
- s t_1 t_2 - \shat t_1 t_2 
- s u_1 u_2 - \shat u_1 u_2 
+ t_1 t_2 u_1 + t_1 t_2 u_2)
\nonumber \\
&-&4 {\rm Im}(H_L^tF_L^{t*})\eps(\eminusi,\eplus,\eminusf,\neubar)
(2 s \shat - 2 t_1 t_2 + u_1^2 + u_2^2 + s u_1 + \shat u_1 + su_2 + \shat u_2).
\label{eq:KLt}
\end{eqnarray}

$K_R^t$ can be obtained from Eq.~(\ref{eq:KLt}) by the substitutions 
$u_1 \leftrightarrow s$ and $u_2 \leftrightarrow \shat$, with an 
additional minus sign for the $G^tF^{t*}$ and $H^tF^{t*}$ terms: 
\begin{eqnarray}
K_R^t&=&4|G_R^t|^2s\shat
\nonumber \\
&+&|H_R^t|^2(s\shat-t_1 t_2 +u_1 u_2+su_1+\shat u_2)
(s\shat-t_1 t_2 +u_1 u_2+su_2+\shat u_1) 
\nonumber \\
&+&|F_R^t|^2(-s^2\shat^2-t_1^2 t_2^2 -u_1^2 u_2^2 
+s^2 t_1 t_2 +\shat^2 t_1 t_2 
+2 s \shat u_1 u_2 +2 t_1 t_2 u_1 u_2 )
\nonumber \\
&+&2 {\rm Re}(G_R^tH_R^{t*})(s^2 \shat  + s \shat^2
+ 2 s \shat u_1 + 2 s \shat u_2 
-s t_1 t_2 -\shat t_1 t_2  
+ s u_1 u_2 + \shat u_1 u_2 ) 
\nonumber \\
&+&8 {\rm Im}(G_R^tH_R^{t*})\eps(\eminusi,\eplus,\eminusf,\neubar)(s-\shat)
\nonumber \\
&-&2 {\rm Re}(G_R^tF_R^{t*})(s^2 \shat -s \shat^2 -s u_1 u_2  + \shat u_1 u_2 
+s t_1 t_2 - \shat t_1 t_2)
\nonumber \\
&+&8 {\rm Im}(G_R^tF_R^{t*})\eps(\eminusi,\eplus,\eminusf,\neubar)(s+\shat)
\nonumber \\
&+& {\rm Re}(H_R^tF_R^{t*})
(s-\shat)(-s^2 \shat - s \shat^2 + u_1^2 u_2 + u_1 u_2 ^2 
- s \shat u_1 - s \shat u_2 
\nonumber \\
&&
+ s t_1 t_2 + \shat t_1 t_2
+ s u_1 u_2 + \shat u_1 u_2 
-  t_1 t_2 u_1 - t_1 t_2 u_2)  
\nonumber \\
&+&4 {\rm Im}(H_R^tF_R^{t*})\eps(\eminusi,\eplus,\eminusf,\neubar)
(s^2 + \shat^2 - 2 t_1 t_2 + 2 u_1 u_2 + s u_1 + \shat u_1 + su_2 + \shat u_2).
\end{eqnarray}

$K^s_L$ can be obtained from Eq.~(\ref{eq:KLt}) by
the substitutions $t_1 \leftrightarrow s$ and
$t_2 \leftrightarrow \shat$, with an additional 
minus sign for the ${\rm Im}(G^sH^{s*})$, ${\rm Im}(G^sF^{s*})$
and ${\rm Im}(F^sH^{s*})$ terms:
\begin{eqnarray}
K_L^s&=&4|G_L^s|^2u_1u_2
\nonumber \\
&+&|H_L^s|^2(-s\shat+t_1 t_2 +u_1 u_2+t_1 u_1+t_2 u_2)
(-s\shat+t_1 t_2 +u_1 u_2+t_1 u_2+t_2 u_1) 
\nonumber \\
&+&|F_L^s|^2(-s^2\shat^2-t_1^2 t_2^2 -u_1^2 u_2^2 
+2 s \shat t_1 t_2 +s \shat  u_1^2 +s \shat u_2^2  + 2 t_1 t_2  u_1 u_2 )
\nonumber \\
&+&2 {\rm Re}(G_L^sH_L^{s*})(u_1^2 u_2 + u_1 u_2^2  
-s \shat  u_1 -s \shat  u_2 
+ t_1 t_2  u_1 + t_1 t_2  u_2 + 2 t_1  u_1 u_2 + 2 t_2  u_1 u_2 
)
\nonumber \\ 
&-&8 {\rm Im}(G_L^sH_L^{s*})\eps(\eminusi,\eplus,\eminusf,\neubar)(u_1-u_2)
\nonumber \\
&+&2 {\rm Re}(G_L^sF_L^{s*})(u_1^2 u_2 -u_1 u_2^2 
+s \shat u_1 - s \shat  u_2 
-t_1 t_2  u_1 + t_1 t_2  u_2 
)
\nonumber \\
&+&8 {\rm Im}(G_L^sF_L^{s*})\eps(\eminusi,\eplus,\eminusf,\neubar)(u_1+u_2)
\nonumber \\
&-& {\rm Re}(H_L^sF_L^{s*})
(u_1-u_2)(t_1^2 t_2 + t_1 t_2^2 -u_1^2 u_2 - u_1 u_2^2 
- s \shat t_1 - s \shat t_2 
+ s \shat  u_1 + s \shat u_2
\nonumber \\
&&
+ t_1 t_2  u_1 + t_1 t_2  u_2 
- t_1  u_1 u_2 - t_2 u_1 u_2 
)
\nonumber \\
&+&4 {\rm Im}(H_L^sF_L^{s*})\eps(\eminusi,\eplus,\eminusf,\neubar)
(-2 s \shat + 2 t_1 t_2 + u_1^2 + u_2^2 + t_1 u_1 
+ t_2  u_1 + t_1 u_2 + t_2  u_2).
\end{eqnarray}

$K_R^s$ can be obtained from Eq.~(\ref{eq:KLt}) by the substitutions
$u_1 \leftrightarrow t_1$ and $u_2 \leftrightarrow t_2$,
with an additional minus sign for the ${\rm Im}(G^sH^{s*})$, 
${\rm Re}(G^sF^{s*})$, and ${\rm Re}(H^sF^{s*})$ terms:
\begin{eqnarray}
K_R^s&=&4|G_R^s|^2t_1 t_2 
\nonumber \\
&+&|H_R^s|^2
(-s\shat+t_1 t_2 +u_1 u_2+t_1 u_1+t_2 u_2)
(-s\shat+t_1 t_2 +u_1 u_2+t_1 u_2+t_2 u_1) 
\nonumber \\
&+&|F_R^s|^2(-s^2\shat^2-t_1^2 t_2^2 -u_1^2 u_2^2 
+s \shat t_1^2 +s \shat t_2^2 +2 s \shat  u_1 u_2 
+2 t_1 t_2  u_1 u_2 )
\nonumber \\
&+&2 {\rm Re}(G_R^sH_R^{s*})(t_1^2 t_2  + t_1 t_2^2
-s \shat t_1 -s \shat t_2  
+ 2 t_1 t_2 u_1 + 2 t_1 t_2 u_2 
+ t_1 u_1 u_2 + t_2 u_1 u_2 ) 
\nonumber \\
&-&8 {\rm Im}(G_R^sH_R^{s*})\eps(\eminusi,\eplus,\eminusf,\neubar)(t_1-t_2)
\nonumber \\
&-&2 {\rm Re}(G_R^sF_R^{s*})(t_1^2 t_2 -t_1 t_2^2 
+s \shat t_1 - s \shat t_2
-t_1 u_1 u_2  + t_2 u_1 u_2 
)
\nonumber \\
&-&8 {\rm Im}(G_R^sF_R^{s*})\eps(\eminusi,\eplus,\eminusf,\neubar)(t_1 + t_2 )
\nonumber \\
&+& {\rm Re}(H_R^sF_R^{s*})
(t_1-t_2 )(-t_1^2 t_2 - t_1 t_2^2 + u_1^2 u_2 + u_1 u_2 ^2 
- s \shat u_1 - s \shat u_2
+ s \shat t_1+ s \shat t_2
\nonumber \\
&&
- t_1 t_2 u_1 - t_1 t_2 u_2 
+ t_1 u_1 u_2 + t_2 u_1 u_2 
)  
\nonumber \\
&-&4 {\rm Im}(H_R^sF_R^{s*})\eps(\eminusi,\eplus,\eminusf,\neubar)
(t_1^2 + t_2 ^2 - 2 s \shat + 2 u_1 u_2 + t_1 u_1 
+ t_2  u_1 + t_1 u_2 + t_2  u_2).
\end{eqnarray}

The interference term $K^{st}$ between the $s$- and $t$-channel diagrams is 
given by:
\begin{eqnarray}
K^{st}&=&-8 {\rm Re}(G_L^tG_L^{s*})u_1 u_2 
\nonumber \\
&-&2 {\rm Re}(G_L^tH_L^{s*})
(u_1^2 u_2 + u_1 u_2^2 - s \shat u_1 - s \shat u_2 
+ t_1 t_2 u_1 + t_1 t_2 u_2 
+ 2 t_1 u_1 u_2 + 2 t_2 u_1 u_2 )
\nonumber \\
&+&8 {\rm Im}(G_L^tH_L^{s*})\eps(\eminusi,\eplus,\eminusf,\neubar)(u_1 - u_2)
\nonumber \\
&-&2 {\rm Re}(G_L^tF_L^{s*})
(u_1^2 u_2 - u_1 u_2^2 + s \shat u_1 - s \shat u_2 
- t_1 t_2 u_1 + t_1 t_2 u_2)
\nonumber \\
&-&8 {\rm Im}(G_L^tF_L^{s*})\eps(\eminusi,\eplus,\eminusf,\neubar)(u_1 + u_2)
\nonumber \\
&-&2 {\rm Re}(H_L^tG_L^{s*})
(u_1^2 u_2 + u_1 u_2^2 
+ s \shat u_1 + s \shat u_2 
+ 2 s u_1 u_2 + 2 \shat u_1 u_2 
- t_1 t_2  u_1 - t_1 t_2  u_2 
)
\nonumber \\
&+&8 {\rm Im}(H_L^tG_L^{s*})\eps(\eminusi,\eplus,\eminusf,\neubar)(u_1 - u_2)
\nonumber \\
&-& {\rm Re}(H_L^tH_L^{s*})
(
- 2 s^2 \shat^2 
- 2 t_1^2 t_2^2 
+ u_1^3 u_2 + u_1 u_2^3 
- s^2 \shat u_1 - s^2 \shat u_2 -s \shat^2  u_1 - s \shat^2 u_2 
\nonumber \\
&&
+ 4 s \shat t_1 t_2
+ s \shat t_1 u_1 + s \shat t_1 u_2 + s \shat t_2 u_1 + s \shat t_2 u_2
- s \shat u_1^2 - s \shat u_2^2  
+ 2 s \shat u_1 u_2 
\nonumber \\
&&
+ s t_1 t_2 u_1 + s t_1 t_2 u_2 + \shat t_1 t_2 u_1 + \shat t_1 t_2 u_2
+ 2 s t_1 u_1 u_2 + 2 s t_2 u_1 u_2 
+ 2 \shat t_1 u_1 u_2 + 2 \shat t_2 u_1 u_2 
\nonumber \\
&&
+ s u_1^2 u_2 + s u_1 u_2^2 + \shat u_1^2 u_2 + \shat u_1 u_2^2 
- t_1^2 t_2 u_1 - t_1^2 t_2 u_2 -t_1 t_2^2  u_1 - t_1 t_2^2 u_2 
- t_1 t_2 u_1^2 - t_1 t_2 u_2^2  
\nonumber \\
&&
+ 2 t_1 t_2 u_1 u_2 
+ t_1 u_1^2 u_2 + t_1 u_1 u_2^2 + t_2  u_1^2 u_2 + t_2 u_1 u_2^2 
)
\nonumber \\
&+&4 {\rm Im}(H_L^tH_L^{s*})\eps(\eminusi,\eplus,\eminusf,\neubar)
(u_1^2 - u_2^2 + s u_1 - s u_2 + \shat u_1 - \shat u_2 
+ t_1 u_1 - t_1 u_2 + t_2 u_1 - t_2 u_2 )
\nonumber \\
&+& {\rm Re}(H_L^tF_L^{s*})
(
- s^2 \shat u_1 + s^2 \shat u_2 - s \shat^2 u_1 + s \shat^2 u_2
- 2 s \shat u_1^2 + 2 s \shat u_2^2  
\nonumber \\
&& + s t_1 t_2 u_1 - s t_1 t_2 u_2 
+ \shat t_1 t_2 u_1 - \shat t_1 t_2 u_2 
- s u_1^2 u_2 + s u_1 u_2^2 
- \shat u_1^2 u_2 + \shat u_1 u_2^2 
)
\nonumber \\
&-&4 {\rm Im}(H_L^tF_L^{s*})\eps(\eminusi,\eplus,\eminusf,\neubar)
(2 s \shat  - 2 t_1 t_2 + 2 u_1 u_2 
+ s u_1 + s u_2 + \shat u_1 + \shat u_2 )
\nonumber \\
&-&2 {\rm Re}(F_L^tG_L^{s*})
(u_1^2 u_2 - u_1 u_2^2 
- s \shat u_1 + s \shat u_2 + t_1 t_2 u_1 - t_1 t_2 u_2)
\nonumber \\
&-&8 {\rm Im}(F_L^tG_L^{s*})\eps(\eminusi,\eplus,\eminusf,\neubar)
(u_1 + u_2)
\nonumber \\
&+& {\rm Re}(F_L^tH_L^{s*})
(
s \shat t_1 u_1 - s \shat t_1 u_2 + s \shat t_2 u_1 - s \shat t_2 u_2 
- t_1^2 t_2 u_1 + t_1^2 t_2 u_2 - t_1 t_2^2 u_1 + t_1 t_2^2 u_2 
\nonumber \\
&&
-2 t_1 t_2 u_1^2 + 2 t_1 t_2 u_2^2 
- t_1 u_1^2 u_2 + t_1 u_1 u_2^2  - t_2 u_1^2 u_2 + t_2 u_1 u_2^2 )
\nonumber \\
&-&4 {\rm Im}(F_L^tH_L^{s*})\eps(\eminusi,\eplus,\eminusf,\neubar)
(-2 s \shat + 2 t_1 t_2 + 2 u_1 u_2 
+ t_1 u_1 + t_1 u_2 + t_2 u_1 + t_2 u_2)
\nonumber \\
&-& {\rm Re}(F_L^tF_L^{s*})
(
2 s^2 \shat^2+ 2 t_1^2 t_2^2 + u_1^3 u_2 + u_1 u_2^3 
-4 s \shat t_1 t_2 
-2 s \shat u_1 u_2 - s \shat u_1^2 - s \shat u_2^2 
-2 t_1 t_2 u_1 u_2 - t_1 t_2 u_1^2 -t_1 t_2 u_2^2)
\nonumber \\
&-&4 {\rm Im}(F_L^tF_L^{s*})\eps(\eminusi,\eplus,\eminusf,\neubar)
(u_1^2 - u_2^2).
\end{eqnarray}
Notice that under the substitution 
$s\leftrightarrow t_1$, $\shat \leftrightarrow t_2$, 
the terms involving $G_L^tG_L^{s*}$, $F_L^tF_L^{s*}$ and $H_L^tH_L^{s*}$ 
are invariant, while the terms involving 
$G_L^tH_L^{s*}$, $G_L^tF_L^{s*}$ and $H_L^tF_L^{s*}$ 
are exchanged with the terms involving 
$H_L^tG_L^{s*}$, $F_L^tG_L^{s*}$ and $F_L^tH_L^{s*}$.

\section{Derivation of the $W^+H^+$ and $G^+H^+$ mixing contribution}
\label{sec:sum}

$W^+H^+$ and $G^+H^+$ mixing contributes through diagrams 
in which the $W^+H^+$ or $G^+H^+$ mixing is attached to the internal 
gauge boson or the  external fermion legs. 
Diagrams with $G^+H^+$ mixing attached to the 
external fermion legs can be neglected since they are proportional to the 
electron or neutrino mass. 
The real part of the $W^+H^+$ and $G^+H^+$ mixing does not contribute 
because of the renormalization condition, Eq.~(\ref{eq:rencond}), and
the Slavnov-Taylor identity, Eq.~(\ref{eq:ST}). 
The calculation of the imaginary part, which is a sum of all the diagrams
involving $W^+H^+$ and $G^+H^+$ mixing, can be simplified as follows.

In unitary gauge, the $G^+H^+$ mixing does not contribute because
$G^+$ is not a physical degree of freedom.
Focusing on the $W$ boson propagator and the attached $W^+H^+$ mixing,
the total $W^+H^+$ mixing contribution can be written as:
\begin{equation}
\left[\frac{-i}{(k_1+k_2)^2-m_W^2}
\left(g^{\mu\nu}-\frac{(k_1+k_2)^{\mu}(k_1+k_2)^{\nu}}{m_W^2}\right)\right]
\cdot
(k_1+k_2)_{\mu}
\longrightarrow
\left(1-\frac{m_{H^{\pm}}^2}{m_W^2}\right)\left[
\frac{-ig^{\mu\nu}}{(k_1+k_2)^2-m_W^2}\cdot (k_1+k_2)_{\mu}\right].
\end{equation}
Here the $(k_1+k_2)_{\mu}$ factor on the left-hand side 
comes from the fact that the $W^+H^+$ mixing 
is given by $-ik_{\mu}\Sigma_{W^+H^+}(k^2)$, where
$k=k_1+k_2$ and $k^2=m_{H^{\pm}}^2$ for on-shell $H^{\pm}$.
On the right-hand side, the term inside the square brackets 
is exactly the same 
as the {\it total} $W^+H^+$ mixing contribution in Feynman gauge.
Therefore, we have
\begin{equation}
\sum(W^+H^+ {\rm contribution})_{\rm unitary\ gauge}
=\left(1-\frac{m_{H^{\pm}}^2}{m_W^2}\right)
\sum(W^+H^+ {\rm contribution})_{\rm Feynman\ gauge}.
\end{equation}
However, due to gauge invariance we can write, 
\begin{equation}
\sum(W^+H^+ {\rm contribution})_{\rm unitary\ gauge}
=\sum(W^+H^+ {\rm contribution})_{\rm Feynman\ gauge}
+(G^+H^+ {\rm contribution})_{\rm Feynman\ gauge}.
\end{equation}
Therefore, 
\begin{equation}
-\frac{m_{H^{\pm}}^2}{m_W^2}
\sum(W^+H^+ {\rm contribution})_{\rm Feynman\ gauge}
=(G^+H^+ {\rm contribution})_{\rm Feynman\ gauge}.
\end{equation}

The total contribution can now be written in terms of the contribution
from $G^+H^+$ mixing, which is easy to calculate as there is only 
one diagram that contributes:
\begin{equation}
{\rm total}=\left(1-\frac{m_W^2}{m_{H^{\pm}}^2}\right)
(G^+H^+ {\rm contribution})_{\rm Feynman\ gauge}.
\end{equation}
The $G^+H^+$ mixing attached to the internal gauge boson line 
always has the same structure
as the effective $W^{+\mu}V^{\nu}H^-$ vertex (for both $s$- and $t$- channel),
and gives rise to an effective contribution to the form-factor $G_V$:
\begin{equation}
i G_V^{\rm eff}=\left(1-\frac{m_W^2}{m_{H^{\pm}}^2}\right)(i g_{VWG} )
\frac{i}{m_{H^{\pm}}^2-m_W^2}
i \hat \Sigma_{G^+H^+}(m_{H^{\pm}}^2),
\end{equation}
where $ g_{VWG} = m_W g_V^G$ and $\hat \Sigma_{G^+H^+}(m_{H^{\pm}}^2) = 
i {\rm Im} \Sigma_{G^+H^+}(m_{H^{\pm}}^2)$.
Using the relation that 
${\rm Im}\Sigma_{G^+H^+}(m_{H^{\pm}}^2)=(m_{H^{\pm}}^2/m_W)
{\rm Im}\Sigma_{W^+H^+}(m_{H^{\pm}}^2)$ from Eq.~(\ref{eq:ST}), we obtain
\begin{equation}
G_V^{\rm eff}=-g_V^G i {\rm Im}\Sigma_{W^+H^+}(m_{H^{\pm}}^2).
\end{equation}
Combining this together with the contribution from the vertex counterterm, 
which can be expressed as $-g_V^G {\rm Re}\Sigma_{W^+H^+}(m_{H^{\pm}}^2)$,
we obtain the result for $G_V^{\rm tot}$ given in Eq.~(\ref{eq:sum}):
\begin{equation}
G_V^{\rm tot}=G_V^{\rm loop}-g_V^G \Sigma_{W^+H^+}(m_{H^{\pm}}^2).
\end{equation}

\end{widetext}


\end{document}